# On Geometric Upper Bounds for Positioning Algorithms in Wireless Sensor Networks

Mohammad Reza Gholami, *Student Member, IEEE*, Erik G. Ström, *Senior Member, IEEE*, Henk Wymeersch, *Member, IEEE*, and Mats Rydström

### Abstract

This paper studies the possibility of upper bounding the position error of an estimate for range based positioning algorithms in wireless sensor networks. In this study, we argue that in certain situations when the measured distances between sensor nodes are positively biased, e.g., in non-line-of-sight conditions, the target node is confined to a closed bounded convex set (a feasible set) which can be derived from the measurements. Then, we formulate two classes of geometric upper bounds with respect to the feasible set. If an estimate is available, either feasible or infeasible, the worst-case position error can be defined as the maximum distance between the estimate and any point in the feasible set (the first bound). Alternatively, if an estimate given by a positioning algorithm is always feasible, we propose to get the maximum length of the feasible set as the worst-case position error (the second bound). These bounds are formulated as nonconvex optimization problems. To progress, we relax the nonconvex problems and obtain convex problems, which can be efficiently solved. Simulation results indicate that the proposed bounds are reasonably tight in many situations.

Authors are with the Division of Communication Systems, Information Theory, and Antennas, Department of Signals and Systems, Chalmers University of Technology, SE-412 96 Gothenburg, Sweden (e-mail: {moreza, erik.strom, henkw, mats.rydstrom}@chalmers.se).

This work was supported by the Swedish Research Council (contract no. 2007-6363).

# I. INTRODUCTION

Recent advances in technology have instigated the use of tiny devices as sensors in large distributed wireless sensor networks (WSNs). A sensor device is capable to sense its environment for monitoring, controlling, or tracking purposes for both civil and military applications [1]. Due to drawbacks in using GPS for WSNs, extracting the position information from the network, also called localization, has been extensively studied in the literature [1]–[6]. It is commonly assumed that there are a number of fixed reference sensors, also called anchors, whose positions are *a priori* known, e.g., by using GPS receivers [7]. To find the position of other sensor nodes at unknown positions, henceforth called target nodes, it is assumed that there are some types of measurements, e.g., time-of-arrival, angle-of-arrival, or received signal strength, taken between sensor nodes [1].

During the last decades, various positioning algorithms have been proposed in the literature. Different positioning approaches can be categorized based on various factors [8]. For instance, as long as an accurate model of measurements and the statistics of the measurement errors are known, classic estimators, e.g., the maximum likelihood (ML) and the least squares (LS) approaches, can be employed successfully to solve the positioning problem. When the distribution of the measurement errors is unknown or the computational complexity of classic estimators is too high, a number of simple techniques can be applied to the problem. For example, based on a geometric interpretation, the authors of [9], [10] formulated the positioning problem as a convex feasibility problem (CFP) and applied the well-known orthogonal projection onto convex sets (POCS) approach to solve the problem. This method turns out to be robust against non-line-of-sight (NLOS) conditions [11]. POCS was previously studied for the CFP and has found applications in several research fields [12], [13].

Positioning algorithms can be evaluated based on different performance metrics such as complexity, accuracy, and coverage [8]. In the literature one way to assess the positioning algorithms is to evaluate the position error, defined as the Euclidian norm of the difference between the position estimate and the true position. There are a number of techniques to evaluate the performance of an algorithm based on the position error. For instance, a lower bound on the mean square position error is a common metric. There exist a number of such lower bounds in the literature, e.g., the Cramér-Rao lower bound (CRLB), which can serve as benchmarks. The CRLB, which gives a lower bound on the variance of any unbiased estimator, can be computed if the probability density function (PDF) of the measurement error is known and satisfies some regularity conditions [14]. Generally, different benchmarks in the literature are used to *statistically* assess a positioning algorithm, which implies that the error in a single position estimate cannot be characterized in a deterministic fashion.

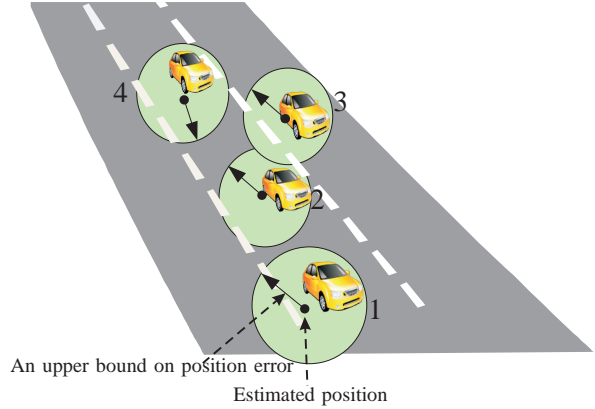

Fig. 1. An example of the application of an upper bound on the position error for traffic safety. A solid circle defines the area in which a vehicle definitely lies. In this figure based on an upper bound on the position error, car 2 and 3 might collide.

Besides a lower bound on the position error, in some applications it may be useful to know the worst-case behavior of the position error. Such knowledge may be useful not only for evaluation of different services provided by WSNs but also for design and resource management [1], [15]. Similarly in evaluation of the worst-case position error, we may be interested in assessing a single point estimate. As an example consider Fig. 1, which shows how a nontrivial (i.e., finite) upper bound on position error can be used by a traffic safety application. If an estimate of a vehicle and a nontrivial upper bound on the position error are available, we can define an area in which the vehicle is certainly located, e.g., a disc centered at the position estimate and with a radius equal to the upper bound on the position error. By this approach, we may be able to decrease the number of collisions between vehicles. In general, computing the maximum possible position error might be difficult, but one may be able to derive an upper bound on the maximum possible position error. To the best of our knowledge, there is no specific work in the literature on deriving an upper bound on the position error. In this study, we aim at tackling this subject.

In general, the concept of an upper bound on the position error (or any estimation error) seems to be shaky. In fact, it is not clear that is meaningful to study upper bounds, since the position error can, in general, be arbitrarily large. In this study, however, we argue that in some situations it is possible to reasonably determine the worst-case position error. For instance, if a target node position belongs to a closed bounded set (a feasible set), the worst-case position error can be defined with respect to the feasible set. For example, for distance-based positioning, if measurement errors are assumed to be positive, a convex set including the target node can be defined from measurements. The feasible set, in which the target node is located, is the intersection of a number of balls (in a 3-dimensional network) or discs (in a 2-dimensional network) centered at the position of reference nodes [16]. The assumption of positively biased measurement errors is fulfilled in some scenarios. For instance, in NLOS conditions, the measured distances are often much larger than the actual distances. Assuming a closed bounded (compact) convex set derived from positively biased distance measurements, a position estimate given by an algorithm can be either feasible or infeasible with respect to the feasible set. If an estimate is available (feasible or infeasible), it is reasonable to define the maximum distance from the estimate to any point in the feasible region as the worst-case position error. This idea yields an upper bound on the position error as the solution of a nonconvex optimization problem. Alternatively, a number of positioning algorithms, e.g., POCS, give one feasible point as an estimate. In this type of estimators, we can upper bound the position error as the maximum length[1] of the feasible set. To find the maximum length of the feasible region, we consider an outer-approximation of the feasible set and find the minimum Euclidean ball or the minimum $\ell_\infty$ ball (minimum bounding box) covering the set. We further relax the nonconvex optimization problem and derive a convex optimization problem. Obviously, if a feasible point is available, the first upper bound, i.e., the maximum distance from the estimate to any point in the feasible region, gives a tighter upper bound compared to the second bound, i.e., the maximum length of the feasible region.

Note that the technique introduced in this paper can be applied to every estimation problem when the unknown parameter vector belongs to a compact, finite-volume, convex set.

---

[1] By the maximum length of a set, we mean the maximum $\ell_2$ norm of the difference between two points (not necessarily a unique pair of points) in the set.

In summary, the main contributions of this study are:

- introducing the concept of an instantaneous upper bound for a single point position estimate when the distance measurements are positively biased, e.g., in NLOS conditions;

- proposing an upper bound on the position error based on a convex relaxation technique when an estimate of the target position is available (feasible or infeasible);

- proposing three upper bounds for an estimator always giving a feasible point as an estimate (e.g., the POCS estimate) based on the idea of the maximum length of the feasible set or a relaxed feasible set including the target node.

The remainder of the paper is organized as follows. Some preliminary requirements are studied in Section II. Section III explains the signal model considered in this paper. In Section IV, a geometric positioning algorithm (POCS) is briefly studied. Two types of upper bounds are derived in Section V. Simulation results are discussed in Section VI. Finally, Section VII makes come concluding remarks.

## II. PRELIMINARIES

### A. Notation

The following notations are used in this study. Lowercase and bold lowercase letters denote scalar values and vectors, respectively. Matrices are written using bold uppercase letters. By $\mathbf{0}_{n \times n}$ we denote the $n$ by $n$ zero matrix, and we use $\mathbf{0}_n$ as the $n$-vector of $n$ zeros. $\mathbf{1}_n$ and $\mathbf{I}_n$ denote the vector of $n$ ones and the $n$ by $n$ identity matrix, respectively. The operator $\mathrm{tr}(\cdot)$ is used to denote the trace of a square matrix. The $\ell_p$ norm is denoted by $\| \cdot \|_p$. Given two matrices $\mathbf{A}$ and $\mathbf{B}$, $\mathbf{A} \succ (\succeq)\mathbf{B}$ means that $\mathbf{A} - \mathbf{B}$ is positive (semi)definite. $\mathbb{S}^n$, $\mathbb{R}^n$, and $\mathbb{R}^n_+$ denote the set of all $n \times n$ symmetric matrices, the set of all $n \times 1$ vectors with real values, and the set of all $n \times 1$ vectors with nonnegative real values, respectively.

### B. Quadratically constrained quadratic programming

Let us consider a quadratically constrained quadratic program (QCQP) as

$$\underset{\mathbf{x} \in \mathbb{R}^n}{\text{maximize}} \ \ \mathbf{x}^T \mathbf{A}_0 \mathbf{x} + 2\mathbf{b}_0^T \mathbf{x} + c_0$$

$$\text{subject to } \mathbf{x}^T \mathbf{A}_i \mathbf{x} + 2\mathbf{b}_i^T \mathbf{x} + c_i \leq 0, \quad i = 1, \dots, N, \tag{1}$$

for $\mathbf{A}_i \in \mathbb{S}^n$, $\mathbf{b}_i \in \mathbb{R}^n$, and $c_i \in \mathbb{R}$.

The QCQP problem (1), in general, is nonconvex and difficult to solve except in some specific cases [17]. For the nonconvex case, there are a number of techniques to approximately solve the problem. One powerful approach is the semidefinite relaxation technique [18]–[23]. Considering a property of the trace operator, i.e., $\mathbf{x}^T \mathbf{A}_i \mathbf{x} = \mathrm{tr}(\mathbf{A}_i \mathbf{x} \mathbf{x}^T)$, the QCQP problem in (1) can be written as

$$\underset{\mathbf{x} \in \mathbb{R}^n}{\text{maximize}} \ \ \mathrm{tr}\left(\mathbf{B}_0 \left[\mathbf{x}^T \ 1\right]^T \left[\mathbf{x}^T \ 1\right]\right)$$
$$\text{subject to} \ \ \mathrm{tr}\left(\mathbf{B}_i \left[\mathbf{x}^T \ 1\right]^T \left[\mathbf{x}^T \ 1\right]\right) \leq 0, \qquad i = 1, \ldots, N, \tag{2}$$

where

$$\mathbf{B}_i = \begin{bmatrix} \mathbf{A}_i & \mathbf{b}_i \\ \mathbf{b}_i^T & c_i \end{bmatrix}. \tag{3}$$

Now, by replacing $\mathbf{Z} = \left[\mathbf{x}^T \ 1\right]^T \left[\mathbf{x}^T \ 1\right]$ and noting that $\mathbf{Z}$ is a rank-1 symmetric positive semidefinite matrix, we get an equivalent problem of (2) as

$$\underset{\mathbf{Z} \in \mathbb{S}^{n+1}}{\text{maximize}} \ \ \mathrm{tr}\left(\mathbf{B}_0 \mathbf{Z}\right)$$
$$\text{subject to} \ \ \mathrm{tr}\left(\mathbf{B}_i \mathbf{Z}\right) \leq 0, \quad i = 1, \ldots, N,$$
$$\mathbf{Z} \succeq 0, \ \ \mathbf{Z}(n+1, n+1) = 1, \ \ \mathrm{rank}(\mathbf{Z}) = 1. \tag{4}$$

Due to the nonconvex constraint $\mathrm{rank}(\mathbf{Z}) = 1$, the optimization problem in (4) is still nonconvex. To change it to a convex problem, we drop the rank-1 constraint and obtain a semidefinite programming problem (SDP) as follows:

$$\underset{\mathbf{Z} \in \mathbb{S}^{n+1}}{\text{maximize}} \ \ \mathrm{tr}\left(\mathbf{B}_0 \mathbf{Z}\right)$$
$$\text{subject to} \ \ \mathrm{tr}\left(\mathbf{B}_i \mathbf{Z}\right) \leq 0, \quad i = 1, \ldots, N,$$
$$\mathbf{Z} \succeq 0, \ \ \mathbf{Z}(n+1, n+1) = 1. \tag{5}$$

To refer to the QPCP formulated in (1) throughout this paper, we use $\mathrm{QP}\{\mathbf{A}_i, \mathbf{b}_i, c_i\}_{i=0}^N$. Similarly, to refer to the SDP relaxation derived in (5) originated from QCQP in (1), we use $\mathrm{SDP}\{\mathbf{A}_i, \mathbf{b}_i, c_i\}_{i=0}^N$. For the optimal values of the objective function of the QCQP and the corresponding SDP relaxation in (1) and in (5), we use $\mathrm{v}_{\mathrm{qp}}\{\mathbf{A}_i, \mathbf{b}_i, c_i\}_{i=0}^{N}$ and $\mathrm{v}_{\mathrm{sdp}}\{\mathbf{A}_i, \mathbf{b}_i, c_i\}_{i=0}^{N}$, respectively. By adopting the relaxation, i.e., dropping the rank-1 constraint, we expand the feasible set, therefore, the objective function in (5) is maximized over a larger set than in (1), thus

$$\mathrm{v}_{\mathrm{qp}}\{\mathbf{A}_i, \mathbf{b}_i, c_i\}_{i=0}^{N} \leq \mathrm{v}_{\mathrm{sdp}}\{\mathbf{A}_i, \mathbf{b}_i, c_i\}_{i=0}^{N}. \tag{6}$$

If the rank of matrix $\mathbf{Z}$ for the optimal solution in (5) is one, then, the solution in (5) is equal to the optimal solution in (1). In general, the optimal solution in (5) has rank higher than one, and then a rank-1 approximation can be applied to the optimal solution in (5), e.g., using a method based on singular value decomposition or an approach based on randomization [20]. For details of rank-1 approximation techniques from a higher rank matrix, see, e.g., [20], [23], [24].

Note that using the Lagrange dual approach, a similar problem as the SDP relaxation in (5) can be obtained [18]. We complete this section by a simple and useful property of the quadratic inequality.

*Lemma 2.1:* For a quadratic function $\mathbf{x}^T\mathbf{A}\mathbf{x} + 2\mathbf{b}^T\mathbf{x} + c$, where $\mathbf{A} \in \mathbb{S}^n$, $\mathbf{b} \in \mathbb{R}^n$, and $c \in \mathbb{R}$ , the following statement always holds true:

$$\mathbf{x}^T\mathbf{A}\mathbf{x} + 2\mathbf{b}^T\mathbf{x} + c \geq 0, \ \forall \mathbf{x} \in \mathbb{R}^n \Longleftrightarrow \begin{bmatrix} \mathbf{A} & \mathbf{b} \\ \mathbf{b}^T & c \end{bmatrix} \succeq 0. \tag{7}$$

*Proof:* See [18]. ∎

## C. Bounds on estimation errors given a realization of the measurement vector

Consider an unknown parameter vector $\mathbf{x} \in \mathbb{R}^n$. Regardless if we model $\mathbf{x}$ as random or unknown deterministic, we can define the set of the possible values of $\mathbf{x}$ as

$$\mathcal{X} \triangleq \{\text{possible values of } \mathbf{x}\} \subseteq \mathbb{R}^n$$

Suppose $\mathbf{m}$ is the observed realization of the (random) measurement vector $\mathbb{M}$. Given the event $\mathbb{M} = \mathbf{m}$, the set of possible values of $\mathbf{x}$ changes to

$$\mathcal{X}(\mathbf{m}) \triangleq \{\text{possible values of } \mathbf{x} : \mathbb{M} = \mathbf{m}\} \subseteq \mathcal{X}.$$

The estimate of $\mathbf{x}$, denoted by $\hat{\mathbf{x}}(\mathbf{m}, \mathbf{f}) \in \mathbb{R}^n$, is a function of the observed data $\mathbf{m}$ and some algorithm tuning parameters, e.g., initialization, step size, termination criterion, etc., which are collected in the vector $\mathbf{f}$. The $\mathbf{f}$-vector is chosen, possibly randomly, from the set $\mathcal{F}$. In other words, $\mathbf{f} \in \mathcal{F}$ completely determines how the estimator maps the observed data $\mathbf{m}$ to the estimate $\hat{\mathbf{x}}$, and the set $\mathcal{F}$ defines a class of estimators. We can now define the set of possible values of $\hat{\mathbf{x}}(\mathbf{m}, \mathbf{f})$ when $\mathbf{f}$ can take on any value in $\mathcal{F}$ as

$$\hat{\mathcal{X}}(\mathbf{m}) \triangleq \{\hat{\mathbf{x}}(\mathbf{m}, \mathbf{f}) : \mathbf{f} \in \mathcal{F}\} \subset \mathbb{R}^n.$$

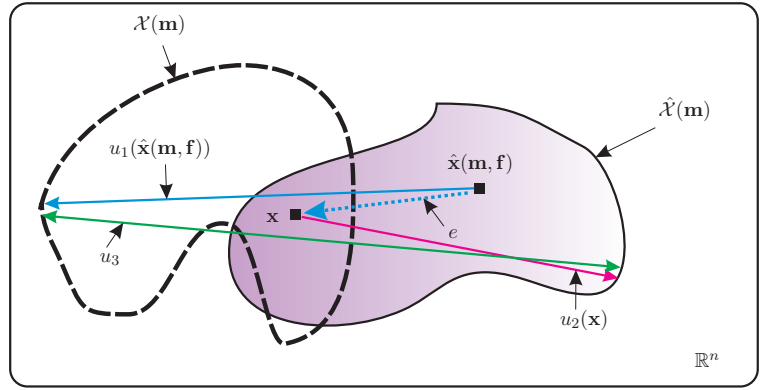

Fig. 2.   Different upper bounds.

We can define three upper bounds on the $\ell_2$ norm of estimation error $e \triangleq \|\hat{\mathbf{x}}(\mathbf{m}, \mathbf{f}) - \mathbf{x}\|_2$ as

$$e \leq u_1(\hat{\mathbf{x}}(\mathbf{m}, \mathbf{f})) \triangleq \sup_{\mathbf{x} \in \mathcal{X}(\mathbf{m})} \|\hat{\mathbf{x}}(\mathbf{m}, \mathbf{f}) - \mathbf{x}\|_2, \tag{8}$$

$$e \leq u_2(\mathbf{x}) \triangleq \sup_{\hat{\mathbf{x}} \in \hat{\mathcal{X}}(\mathbf{m})} \|\hat{\mathbf{x}} - \mathbf{x}\|_2, \tag{9}$$

$$e \leq u_3 \triangleq \sup_{\mathbf{x} \in \mathcal{X}(\mathbf{m}), \ \hat{\mathbf{x}} \in \hat{\mathcal{X}}(\mathbf{m})} \|\hat{\mathbf{x}} - \mathbf{x}\|_2. \tag{10}$$

We note that all bounds depends on $\mathbf{m}$, which, for simplicity, is neglected in the notation. Moreover, it is easy to see that $u_1(\hat{\mathbf{x}}(\mathbf{m}, \mathbf{f})) \leq u_3$ and $u_2(\mathbf{x}) \leq u_3$. Fig. 2 graphically shows the different upper bounds.

*Remark 1:* the bound $u_1(\hat{\mathbf{x}}(\mathbf{m}, \mathbf{f}))$ is an upper bound of the norm of the estimation error for a certain estimate ($\mathbf{f}$ and $\mathbf{m}$ are fixed). Hence, if $u_1(\hat{\mathbf{x}}(\mathbf{m}, \mathbf{f}))$ can be computed together with the estimate, this would greatly increase the value of the estimate, since we can now guarantee that the norm of the estimation error in $\hat{\mathbf{x}}(\mathbf{m}, \mathbf{f})$ does not exceed $u_1(\hat{\mathbf{x}}(\mathbf{m}, \mathbf{f}))$. This is a much stronger statement than to provide a statistical quality measure, such as the mean-squared error of the estimator,

$$\mathbb{E}_{\mathbb{M}}\{\|\hat{\mathbf{x}}(\mathbf{m}, \mathbf{f}) - \mathbf{x}\|_2^2\},$$

where $\mathbb{E}_{\mathbb{M}}$ denotes expectation over the distribution of $\mathbb{M}$.

*Remark 2:* the bound $u_3$ could potentially be computed together with the estimate and is therefore of value in a practical situation. However, $u_3$ will only be interesting if it is easier to compute than $u_1(\hat{\mathbf{x}}(\mathbf{m}, \mathbf{f}))$, since $u_1(\hat{\mathbf{x}}(\mathbf{m}, \mathbf{f})) \leq u_3$.

*Remark 3:* the bound $u_2(\mathbf{x})$ can be interpreted as the error of the worst estimate that is computed from the observed data $\mathbf{m}$ by the class of estimators defined by $\mathcal{F}$. This is useful to judge the worst case performance of a class of estimators. However, since the bound is a function of $\mathbf{x}$ (the unknown parameter), it cannot be computed together with an estimate, and its practical value is therefore limited. We can also formulate lower bounds by replacing $\sup$ with $\inf$ in Eqs. (8)– (10),

$$e \geq \ell_1(\hat{\mathbf{x}}(\mathbf{m}, \mathbf{f})) \triangleq \inf_{\mathbf{x} \in \mathcal{X}(\mathbf{m})} \|\hat{\mathbf{x}}(\mathbf{m}, \mathbf{f}) - \mathbf{x}\|_2, \tag{11}$$

$$e \geq \ell_2(\mathbf{x}) \triangleq \inf_{\hat{\mathbf{x}} \in \hat{\mathcal{X}}(\mathbf{m})} \|\hat{\mathbf{x}} - \mathbf{x}\|_2, \tag{12}$$

$$e \geq \ell_3 \triangleq \inf_{\mathbf{x} \in \mathcal{X}(\mathbf{m}), \ \hat{\mathbf{x}} \in \hat{\mathcal{X}}(\mathbf{m})} \|\hat{\mathbf{x}} - \mathbf{x}\|_2. \tag{13}$$

In general, there are no guarantees that any of the bounds in Eqs. (8)–(13) are nontrivial, i.e., that the upper bounds are finite and the lower bounds are greater than zero. For example, if the set $\mathcal{X}(\mathbf{m})$ or $\hat{\mathcal{X}}(\mathbf{m})$ is unbounded, it is clear that the upper bound (8) or (10) is trivial. However, as we will see in the remainder of this paper, there are indeed practical situations when the bounds are nontrivial.

## III. System Model

Let us consider an $n$-dimensional network, $n = 2$ or 3, with $N$ reference nodes at known positions $\mathbf{a}_i = [a_{i,1} \ \cdots \ a_{i,n}]^T \in \mathbb{R}^n$, $i = 1, ..., N$. Suppose that a target node is placed at an unknown position $\mathbf{x} = [x_1 \ \cdots \ x_n]^T \in \mathbb{R}^n$. The range measurement between the target and reference node $i$ is given by

$$\hat{d}_i = d_i(\mathbf{x}, \mathbf{a}_i) + \epsilon_i, \qquad i = 1, \dots, N, \tag{14}$$

where $d_i(\mathbf{x}, \mathbf{a}_i)$ is the actual Euclidian distance between the target node and reference node $i$, i.e., $d_i(\mathbf{x}, \mathbf{a}_i) = \|\mathbf{a}_i - \mathbf{x}\|_2$, and $\epsilon_i$ is the measurement error.

In the literature the measurement error is commonly modeled as a zero mean Gaussian random variable [1], [4], [25]. In some scenarios, however, other distributions seem to be more reasonable. For instance, in NLOS conditions the measured distances are larger than the actual distances with high probability. A number of distributions have been considered to model NLOS conditions, e.g., an exponential distribution or a uniform distribution [26]. The Gaussian distribution with large positive mean has also been considered to model the NLOS condition [26], [27]. In this paper for the purpose of deriving the upper bound, we assume that the distance measurements are positively biased, meaning the measurement errors are nonnegative. The positive measurement assumption can be fulfilled, e.g., in NLOS conditions (with high probability).

The positioning problem, then, is to find the position of the target node based on the positions of $N$ reference nodes and measurements made in (14).

## IV. Positioning algorithms

A classic method to solve the problem of positioning based on measurements taken in (14) is to employ an ML estimator if the distribution of the measurement error $\epsilon_i$ is known. Otherwise, when the statistics of measurement errors are unknown, one can apply the LS minimization as [14], [28]

$$\hat{\mathbf{x}} = \arg \min_{\mathbf{x} \in \mathbb{R}^n} \sum_{i=1}^{N} \left( \hat{d}_i - d_i(\mathbf{x}, \mathbf{a}_i) \right)^2. \tag{15}$$

The solution to (15) coincides with the ML estimate if the measurement errors are zero mean, independent and identically distributed Gaussian random variables [14]. In general, the LS and ML problems are nonconvex and difficult to solve. To avoid difficulty in solving the ML (or LS), authors in [10] took a geometric interpretation into account and formulated the positioning problem as a CFP and applied the well-known POCS approach to solve the positioning problem.

To formulate POCS, note that in the absence of measurement errors, i.e., $\hat{d}_i = d_i(\mathbf{x}, \mathbf{a}_i)$, it is clear that the target, at unknown position $\mathbf{x}$, can be found in the intersection of a number of spheres with radii $d_i(\mathbf{x}, \mathbf{a}_i)$ and centers $\mathbf{a}_i$. For nonnegative measurement errors, we relax spheres to balls and deduce that the target definitely lies inside the intersection of a number of balls. Let us define the (closed bounded) ball $\mathcal{B}_i$ centered at $\mathbf{a}_j$ as

$$\mathcal{B}_i \triangleq \big\{ \mathbf{x} \in \mathbb{R}^n \ : \ \|\mathbf{x} - \mathbf{a}_i\|_2 \leq \hat{d}_i \big\}, \qquad i = 1, \ldots, N. \tag{16}$$

It is then reasonable to define an estimate of $\mathbf{x}$ as a point in the intersection $\mathcal{B}$ (a closed bounded set) of the balls $\mathcal{B}_i$ (a feasible point) as

$$\hat{\mathbf{x}} \in \mathcal{B} \triangleq \bigcap_{i=1}^{N} \mathcal{B}_i. \tag{17}$$

Therefore, the positioning problem can be rendered to the following convex feasibility problem (CFP):

$$\begin{aligned} &\underset{\mathbf{x} \in \mathbb{R}^n}{\text{minimize}}\, 0 \\ &\text{subject to } \|\mathbf{x} - \mathbf{a}_i\| \leq \hat{d}_i, \ \ i = 1, \ldots, N. \end{aligned} \tag{18}$$

To solve (18), we note that CFP can be reformulated by minimizing the following convex function

$$f(\mathbf{x}) \triangleq \max\{\text{dist}(\mathbf{x}, \mathcal{B}_1), \ldots, \text{dist}(\mathbf{x}, \mathcal{B}_N)\}, \tag{19}$$

with $\text{dist}(\mathbf{x}, \mathcal{B}_i)$ denoting the minimum distance between $\mathbf{x}$ and any point in set $\mathcal{B}_i$.

Using negative subgradient updating method [12], [29], we can obtain a solution to (19) by

$$\mathbf{x}^{k+1} = \mathbf{x}^k - \alpha_k \mathbf{g}^k, \quad k = 0, 1, \ldots, \tag{20}$$

where $\mathbf{x}^k$ is the $k$th iterate, $\alpha_k$ is the $k$th step size, and $\mathbf{g}^k$ is a subgradient[2]. A subgradient $\mathbf{g}^k$ of $f$ at $\mathbf{x}^k$ can be computed as

$$\mathbf{g}^k = \begin{cases} 0, & \text{if } f(\mathbf{x}^k) = 0, \\ \dfrac{\mathbf{x}^k - \mathcal{P}_{\mathcal{B}_j}(\mathbf{x}^k)}{\|\mathbf{x}^k - \mathcal{P}_{\mathcal{B}_j}(\mathbf{x}^k)\|_2}, & \text{if } f(\mathbf{x}^k) \neq 0, \ \text{dist}(\mathbf{x}^k, \mathcal{B}_j) \geq \text{dist}(\mathbf{x}^k, \mathcal{B}_i), \ \forall i \neq j, \end{cases} \tag{21}$$

where $\mathcal{P}_{\mathcal{B}_j}(\mathbf{x}^k)$ is the orthogonal projection of $\mathbf{x}^k$ onto the set $\mathcal{B}_j$. By choosing the step size as $\alpha_k = f(\mathbf{x}^k)/\|\mathbf{g}^k\|_2^2$ in (20), according to Polyak approach [12], we derive the following approach, called alternating projections [30] or POCS, for updating

$$\mathbf{x}^{k+1} = \mathcal{P}_{\mathcal{B}_j}(\mathbf{x}^k), \quad k = 0, 1, \ldots, \tag{22}$$

---

[2]Let $\mathcal{D}$ be a nonempty set in $\mathbb{R}^n$. A vector $\mathbf{g} \in \mathbb{R}^n$ is a subgradient of a function $f : \mathcal{D} \to \mathbb{R}$ at $\mathbf{x} \in \mathcal{D}$ if $f(\mathbf{y}) \geq f(\mathbf{x}) + \mathbf{g}^T(\mathbf{y} - \mathbf{x})$ for all $\mathbf{y} \in \mathcal{D}$ [12].

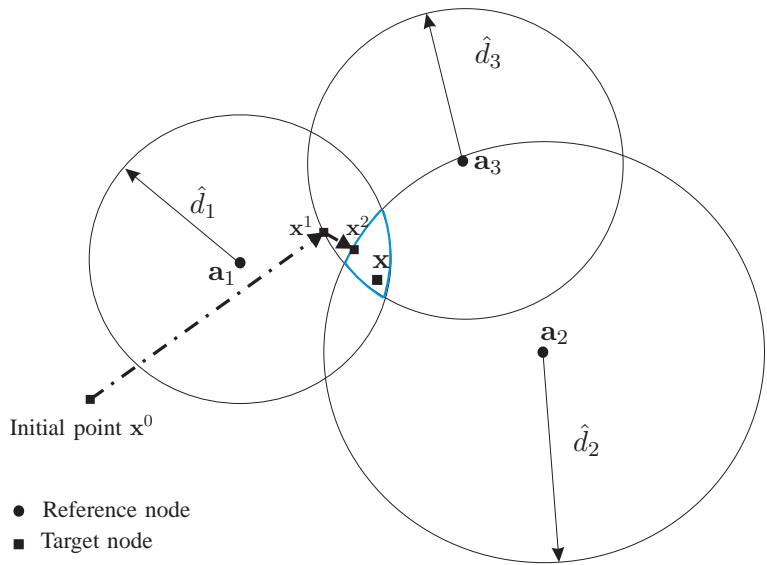

Fig. 3. A 2-dimensional network consisting of three reference nodes and one target node. For nonnegative measurement errors, the target node at position $\mathbf{x}$ is found in the intersection of three discs. The POCS estimate converges to a point $\hat{\mathbf{x}}$ inside the intersection area (in this case on the boundary).

where index $\hat{j}$ is the one used in (21).

As mentioned before, POCS gives an estimate that is feasible (if the intersection $\mathcal{B}$ is nonempty). In each step, POCS projects the current point $\mathbf{x}^k$ onto the farthest convex set. For example, Fig. 3 shows a 2-dimensional network in which the measured distances in reference nodes are positively biased. The POCS' estimate in this figure converges to a point in the intersection of three discs after two iterations. For more details on variations of the POCS algorithm and the application of POCS for the positioning problem, we refer the reader to [12] and [9], [11], [31], respectively.

## V. Geometric upper bounds

In this study, taking the assumption of positively biased measurement errors into account and considering discussions in Section II-C, we derive two different upper bounds. The first bound is derived based on the availability of an estimate. If such an estimate is available (feasible or infeasible), we can bound it by finding the maximum distance between the estimate and any point in the feasible set. The second bound is derived without the need for an estimate, as the maximum length of the intersection set.

Let us define the norm of position estimate, which we call the position error, as

$$e \triangleq \|\hat{\mathbf{x}} - \mathbf{x}\|_2, \tag{23}$$

where $\hat{\mathbf{x}}$ is an estimate of the target node position given by a positioning algorithm. In a practical scenario it is not possible to compute the exact position error in (23) since the position of a target node is unknown. Therefore, we may compute a lower or an upper bound on the position error for evaluation of an estimate. According to discussions in Section II-C, it seems that the plausible definition for the maximum position error, when a single estimate is available, can be considered as

$$e \leq v_{\max,1} \triangleq \max_{\mathbf{x} \in \mathcal{B}} \|\hat{\mathbf{x}} - \mathbf{x}\|_2, \tag{24}$$

where $\mathcal{B}$ defines a set (closed bounded) in which the target node $\mathbf{x}$ belongs. In fact, definition (24) is a special case of the upper bound defined in (8) in Section II-C when $\mathcal{X}(\mathbf{m}) = \mathcal{B}$. In other words, (24) defines the largest distance from a point to a set.

Alternatively, if an algorithm always produces one point in the feasible set $\mathcal{B}$ as an estimate, we are still able to define an upper bound on the position error, even without having access to an estimate, by setting $\mathcal{X}(\mathbf{m}) = \hat{\mathcal{X}}(\mathbf{m}) = \mathcal{B}$ in (10),

$$e \leq v_{\max,3} \triangleq \max_{\mathbf{x},\mathbf{y} \in \mathcal{B}} \|\mathbf{x} - \mathbf{y}\|_2. \tag{25}$$

### A. A bound for the case an estimate exists

As mentioned in previous section, we can upper bound the position error due to an estimate $\hat{\mathbf{x}}$ (either feasible or infeasible), by solving the optimization problem (24). The solution is found on the boundary of set $\mathcal{B}$. For example, let us consider Fig. 4 where an estimate $\hat{\mathbf{x}}$ of the target node position inside the intersection of three discs is available. The position error and the maximum position error are shown in this figure. Instead of directly solving the problem in (24), we consider a QCQP problem $\text{QP}\{\mathbf{A}_i, \mathbf{b}_i, c_i\}_{i=0}^N$, where

$$\mathbf{A}_i = \mathbf{I}_n, \qquad \mathbf{b}_i = \begin{cases} -\hat{\mathbf{x}}, & \text{if } i = 0, \\ -\mathbf{a}_i, & \text{otherwise,} \end{cases} \qquad c_i = \begin{cases} \|\hat{\mathbf{x}}\|^2, & \text{if } i = 0, \\ \|\mathbf{a}_i\|^2 - \hat{d}_i^2, & \text{otherwise.} \end{cases} \tag{26}$$

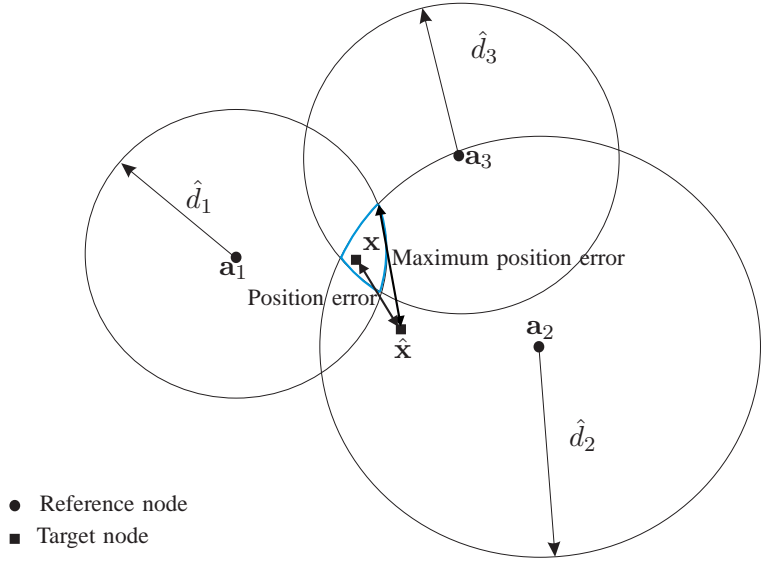

Fig. 4. The position error and the maximum position error for an estimate $\hat{\mathbf{x}}$ of the target for the network considered in Fig. 3.

Obviously, $v_{\mathrm{qp}}\{\mathbf{A}_i, \mathbf{b}_i, c_i\}_{i=0}^{N} = v_{\max,1}^2$. The optimization problem in (24) is nonconvex which makes the problem complicated. To solve the problem, we employ a relaxation technique. Following the procedures explained in Section II-B, we can get a relaxed SDP problem as $\mathrm{SDP}\{\mathbf{A}_i, \mathbf{b}_i, c_i\}_{i=0}^{N}$ and the maximum position error can be upper bounded as

$$e = \|\hat{\mathbf{x}} - \mathbf{x}\|_2 \leq v_{\max,1} \leq \sqrt{v_{\mathrm{sdp}}\{\mathbf{A}_i, \mathbf{b}_i, c_i\}_{i=0}^{N}}. \tag{27}$$

In order to investigate the tightness of the upper-bound derived in (27), we can derive a lower-bound on $v_{\mathrm{qp}}\{\mathbf{A}_i, \mathbf{b}_i, c_i\}_{i=0}^{N}$. Let us write the QCQP problem $\mathrm{QP}\{\mathbf{A}_i, \mathbf{b}_i, c_i\}_{i=0}^{N}$ parameterized in (26) as

$$\underset{\mathbf{x} \in \mathbb{R}^n, \, \tau \in \mathbb{R}}{\text{maximize}} \ \mathrm{tr}\left(\mathbf{B}\left[\mathbf{x}^T \ \tau\right]^T \left[\mathbf{x}^T \ \tau\right]\right)$$

$$\text{subject to} \ \mathrm{tr}\left(\mathbf{B}_i\left[\mathbf{x}^T \ \tau\right]^T \left[\mathbf{x}^T \ \tau\right]\right) \leq t_i, \ \ i = 1, \ldots, N+1, \tag{28}$$

where

$$\mathbf{B}_{N+1} = \begin{bmatrix} \mathbf{0}_{n \times n} & \mathbf{0}_n \\ \mathbf{0}_n^T & 1 \end{bmatrix}, \quad \mathbf{B} = \begin{bmatrix} \mathbf{I}_n & -\hat{\mathbf{x}} \\ -\hat{\mathbf{x}}^T & \|\hat{\mathbf{x}}\|^2 \end{bmatrix}, \quad \mathbf{B}_i = \begin{bmatrix} \mathbf{I}_n & -\mathbf{a}_i \\ -\mathbf{a}_i^T & \|\mathbf{a}_i\|_2^2 + \epsilon^2 \end{bmatrix},$$

$$t_i = \hat{d}_i^2 + \epsilon^2, \ i \leq N, \quad t_{N+1} = 1, \tag{29}$$

where $\epsilon \neq 0$ is any nonzero real value. It is seen that $\mathbf{B}_i \succ 0$ for $1 \leq i \leq N$. Then, $\sum_{i=1}^{N+1} \mathbf{B}_i \succ 0$, meaning the interior of the feasible set is nonempty.

*Proposition 5.1:* A lower bound on the optimal value of $\mathrm{QP}\{\mathbf{A}_i, \mathbf{b}_i, c_i\}_{i=0}^{N}$ parameterized in (26) based on the optimal value $v_{\mathrm{sdp}}\{\mathbf{A}_i, \mathbf{b}_i, c_i\}_{i=0}^{N}$, can be obtained as

$$\sqrt{\alpha \, v_{\mathrm{sdp}}\{\mathbf{A}_i, \mathbf{b}_i, c_i\}_{i=0}^{N}} \leq v_{\mathrm{qp}}\{\mathbf{A}_i, \mathbf{b}_i, c_i\}_{i=0}^{N}, \tag{30}$$

where

$$\alpha = \frac{1}{2\ln(2(N+1)\mu)}, \qquad \mu = \min\{N+1, n+1\}. \tag{31}$$

*Proof:* Recalling the results of [32], which determines a lower bound on the optimal value of a QCQP based on its relaxed SDP, we get a lower bound on the optimal value of (28), which is exactly $v_{\mathrm{qp}}\{\mathbf{A}_i, \mathbf{b}_i, c_i\}_{i=0}^{N}$, as

$$\alpha \, v_{\mathrm{sdp}}\{\mathbf{A}_i, \mathbf{b}_i, c_i\}_{i=0}^{N} \leq v_{\mathrm{qp}}\{\mathbf{A}_i, \mathbf{b}_i, c_i\}_{i=0}^{N}, \tag{32}$$

where

$$\alpha = \frac{1}{2\ln(2(N+1)\mu)}, \quad \mu = \min\{N+1, \max_i \mathrm{rank}(\mathbf{B}_i)\}.$$

It is clear that $\mathrm{rank}(\mathbf{B}_i) = n+1$. Therefore, a lower bound on $v_{\mathrm{qp}}\{\mathbf{A}_i, \mathbf{b}_i, c_i\}_{i=0}^{N}$ can be derived as (30). ∎

For details of deriving lower bounds on a nonconvex QCQP, we refer the reader to [18], [24], [32] and references therein.

### B. Bound regarding the feasible set

In this section, we investigate another upper bound defined in (25) and repeated here for convenience

$$v_{\mathrm{max},3} = \max\left\{\|\mathbf{x} - \mathbf{v}\|_2 : \ \mathbf{x}, \mathbf{v} \in \mathcal{B}\right\}. \tag{33}$$

If a feasible point $\hat{\mathbf{x}} \in \mathcal{B}$ is available, it is expected that the first upper bound $v_{\mathrm{max},1}$ yields a tighter bound compared to the bound defined in (33) (the maximum length of the intersection). In fact for a fixed $\hat{\mathbf{x}} \in \mathcal{B}$,

$$\max_{\mathbf{x}, \mathbf{w} \in \mathcal{B}} \|\mathbf{x} - \mathbf{w}\|_2 \geq \max_{\mathbf{x} \in \mathcal{B}} \|\mathbf{x} - \hat{\mathbf{x}}\|_2. \tag{34}$$

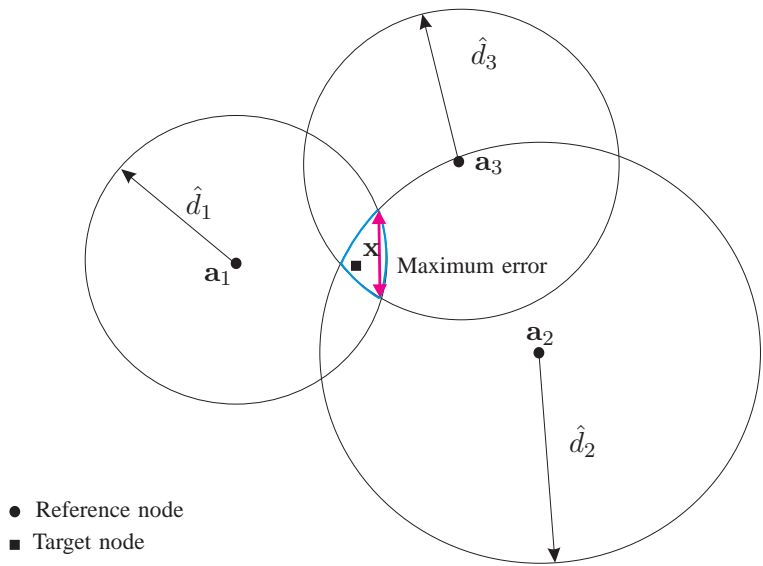

Fig. 5. Maximum Euclidian distance of the intersection as the maximum position error for an estimate inside the intersection area of Fig. 3.

The optimization problem in (33) is nonconvex. Geometrically, it can be imagined as the diameter of the minimum ball enclosing the intersection. Instead of solving the problem formulated in (33), we find a minimum ball covering the intersection $\mathcal{B}$. Let us consider the center $\mathbf{x}_c$ and the radius $R$ of such a ball and formulate the minimum ball enclosing the intersection $\mathcal{B}$ in decision variables $\mathbf{x}_c$ and $\gamma = R^2$ as

$$\underset{\mathbf{x}_c \in \mathbb{R}^n, \; \gamma \in \mathbb{R}_+}{\text{minimize}} \quad \gamma$$

$$\text{subject to } \|\mathbf{x} - \mathbf{x}_c\|^2 \leq \gamma, \; \mathbf{x} \in \mathcal{B}. \tag{35}$$

Let the optimal solution of (35) be $v'_{\max,3}$. Then, $v_{\max,3} = 2\sqrt{v'_{\max,3}}$. Fixing $\mathbf{x}_c$ in (35), using Lemma 2.1, and following a similar approach as used in [33], we can get the following optimization problem to find the minimum ball enclosing the intersection $\mathcal{B}$:

$$\underset{\gamma \in \mathbb{R}_+, \, \boldsymbol{\lambda} \in \mathbb{R}_+^N}{\text{minimize}} \;\; \gamma$$

$$\text{subject to } \begin{bmatrix} (\sum_{i=1}^N \lambda_i - 1)\mathbf{I}_n & \mathbf{x}_c - \sum_{i=1}^N \lambda_i \mathbf{a}_i \\ (\mathbf{x}_c - \sum_{i=1}^N \lambda_i \mathbf{a}_i)^T & \gamma - \|\mathbf{y}\|_2^2 + \sum_{i=1}^N \lambda_i(\|\mathbf{a}_i\|_2^2 - \hat{d}_i^2) \end{bmatrix} \succeq 0,$$

$$\mathbf{x}_c \in \mathbb{R}^n. \tag{36}$$

Taking similar steps as done in [33], which implies for the optimal solution $\sum_{i=1}^N \lambda_i = 1$ and $\mathbf{x}_c = \sum_{i=1}^N \lambda_i \mathbf{a}_i$, we can obtain an optimization problem to find an upper bound on the squared radius of the minimum ball enclosing the set $\mathcal{B}$ in the Euclidian norm sense as

$$\underset{\boldsymbol{\lambda} \in \mathbb{R}_+^N}{\text{minimize}} \; \|\sum_{i=1}^N \lambda_i \mathbf{a}_i\|_2^2 - \sum_{i=1}^N \lambda_i(\|\mathbf{a}_i\|_2^2 - \hat{d}_i^2)$$

$$\text{subject to } \sum_{i=1}^N \lambda_i = 1. \tag{37}$$

Finally, an upper bound on the maximum length of $\mathcal{B}$ is given by

$$v_{\max,3} \leq 2R, \tag{38}$$

where $R = \sqrt{\|\sum_{i=1}^N \lambda_i \mathbf{a}_i\|_2^2 - \sum_{i=1}^N \lambda_i(\|\mathbf{a}_i\|_2^2 - \hat{d}_i^2)}$.

It has been proved in [33] that when the number of constraints $N$ (here the number of reference nodes) is equal or less than $n$ (the size of dimension), (37) gives the optimal solution to (35). Otherwise when $N > n$, the optimal solution in (37) is an upper-bound to the optimal solution in (35). The upper bound obtained by solving (37) then gives the maximum Euclidian length of the intersection.

Another approach to compute an upper bound on $v_{\max,3}$ is to replace $\mathcal{B}$ with an enclosing set in (25). We will in the following consider two such sets. The first enclosing set is the bounding box[3] for $\mathcal{B}$, and, given the bounding box, it is very easy to compute an upper bound on $v_{\max,3}$, see Fig. 6.

The second enclosing set is found be replacing $\mathcal{B}_i$ with their bounding boxes, i.e., the $\ell_2$ balls in (16) are replaced by the corresponding $\ell_\infty$ balls,

$$\mathcal{B}_i' = \{\mathbf{x} \in \mathbb{R}^n : \|\mathbf{x} - \mathbf{a}_i\|_\infty \leq \hat{d}_i\},$$

---

[3]By the bounding box of the set $\mathcal{A}$, we mean the smallest cuboid [34] that is enclosing $\mathcal{A}$.

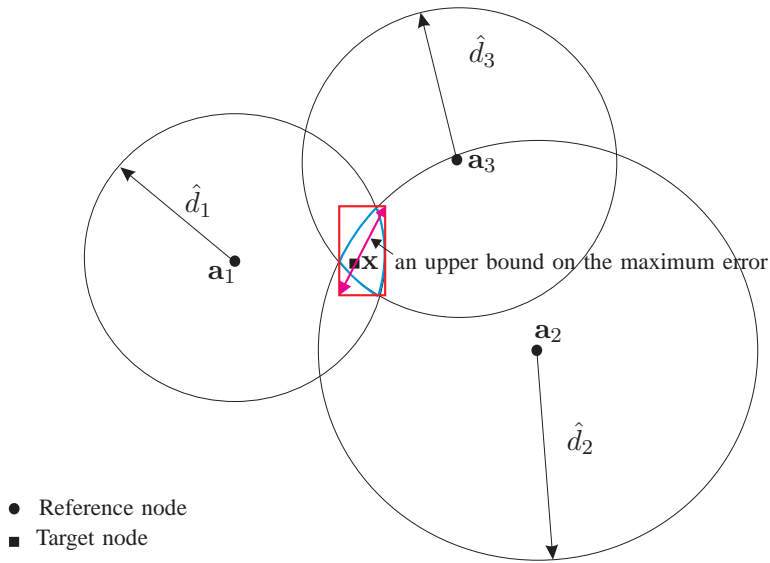

Fig. 6.   The maximum length of the bounding box of the intersection as an upper bound for the network considered in Fig. 3.

and noting that

$$\mathcal{B} \subseteq \mathcal{B}' \triangleq \bigcap_{i=1}^{N} \mathcal{B}_i'.$$

Hence, an upper bound to $v_{\max,3}$ is found by considering the length of $\mathcal{B}'$, see Fig. 7.

To compute the bounding box for $\mathcal{B}$, we study the following optimization problem:

$$\text{maximize } \|\mathbf{x} - \mathbf{y}\|_{\infty}$$

$$\text{subject to } \mathbf{x}, \mathbf{y} \in \mathcal{B}. \tag{39}$$

The optimization problem in (39) again is nonconvex. Using the definition of the $\ell_{\infty}$ norm, we can write

$$\underset{\mathbf{x},\mathbf{y}}{\text{maximize}} \ \ \max(|x_1 - y_1|, \ldots, |x_n - y_n|)$$

$$\text{subject to } \ \ \mathbf{x}, \mathbf{y} \in \mathcal{B}. \tag{40}$$

The $\max$ function in (40) can be computed as

$$\max\{\alpha_1, \ldots, \alpha_n\} = \alpha_i \iff \alpha_i \geq \alpha_j, \ \forall j. \tag{41}$$

Using a dummy variable $\beta$, we have

$$\max\{\alpha_1, \ldots, \alpha_n\} \geq \beta \iff \alpha_1 \geq \beta \text{ or } \alpha_2 \geq \beta \ \ldots \ \text{or } \alpha_n \geq \beta. \tag{42}$$

Thus, using a simple technique, we need to solve two optimization problems for every dimension $\ell$ as follows:

$$\underset{\beta \in \mathbb{R}}{\text{maximize}} \ \ \beta$$

$$\text{subject to} \ \ \|\mathbf{x} - \mathbf{a}_i\| \le \hat{d}_i, \quad i = 1, \ldots, N,$$

$$x_\ell \ge \beta, \tag{43a}$$

$$\underset{\beta \in \mathbb{R}}{\text{minimize}} \ \ \beta$$

$$\text{subject to} \ \ \|\mathbf{x} - \mathbf{a}_i\| \le \hat{d}_i, \quad i = 1, \ldots, N$$

$$x_\ell \le \beta. \tag{43b}$$

The optimization problems in (43) are called the second order cone program which is a special case of the quadratic programming. It can be easily transformed to an SDP [17]. Suppose that the optimal solution to problems (43a) and (43b) along a dimension $\ell$ are $x_{\ell_1}^*$ and $x_{\ell_2}^*$, respectively. Let the maximum length for the $\ell$th dimension be $v_{\text{socp},\ell} = |x_{\ell_1}^* - x_{\ell_2}^*|$. Then, the maximum length of the intersection can be upper bounded as

$$v_{\text{socp}} = \sqrt{\sum_{i=1}^{n} (v_{\text{socp},\ell})^2}. \tag{44}$$

Thus

$$v_{\max,3} \le v_{\text{socp}}. \tag{45}$$

To compute the upper bound on $v_{\max,3}$ based on $\mathcal{B}'$, we consider the following optimization problem:

$$\underset{\mathbf{x}, \mathbf{y}}{\text{maximize}} \ \|\mathbf{x} - \mathbf{y}\|_\infty$$

$$\text{subject to} \ \mathbf{x}, \mathbf{y} \in \mathcal{B}', \tag{46}$$

For example Fig. 7 shows the concept of relaxing the constraint for a 2-dimensional network. Following the same procedure to obtain (43), we obtain two optimization problems, called linear programs (LPs),

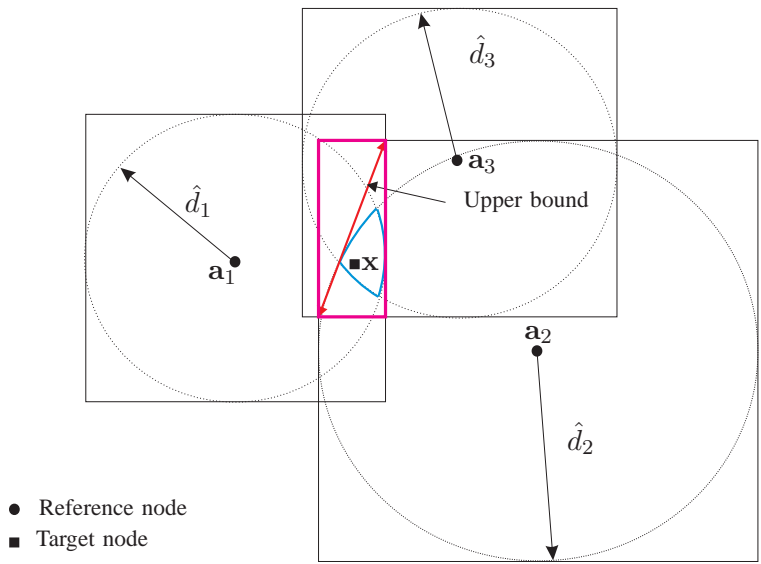

Fig. 7. Every constraint is replaced with a bounding box and then a bounding box enclosing the intersection of relaxed constraints is computed. The maximum length of the bounding box enclosing the intersection gives an upper bound for the network considered in Fig. 3.

for every dimension. For instance, two LPs for the $\ell$th dimension can be written as

$$\underset{t_\ell \in \mathbb{R}}{\text{maximize}} \quad t_\ell$$

$$\text{subject to} \quad t_\ell - a_{i,\ell} - \hat{d}_i \le 0,$$

$$t_\ell - a_{i,\ell} + \hat{d}_i \le 0, \quad i = 1, \dots, N, \tag{47a}$$

$$\underset{t_\ell \in \mathbb{R}}{\text{minimize}} \quad t_\ell$$

$$\text{subject to} \quad t_\ell - a_{i,\ell} - \hat{d}_i \le 0,$$

$$a_{i,\ell} - t_\ell + \hat{d}_i \le 0, \quad i = 1, \dots, N. \tag{47b}$$

The optimal solution to the optimization problem (47), i.e., $t^*_{\ell_1}$ and $t^*_{\ell_2}$, are simply computed as

$$t^*_{\ell_1} = \min\{a_{1,\ell} + \hat{d}_1, \dots, a_{N,\ell} + \hat{d}_N\}, \quad t^*_{\ell_2} = \max\{a_{1,\ell} - \hat{d}_1, \dots, a_{N,\ell} - \hat{d}_N\}. \tag{48}$$

Let $v_{\text{lp},\ell} = |t^*_{\ell_1} - t^*_{\ell_2}|, \ \ell = 1, \dots, n$, be the maximum length along the $\ell$th dimension. The maximum

SUMMARY OF BOUNDS.

| Definition: | Eqn. |
|---|---|
| $e \triangleq \|\hat{\mathbf{x}} - \mathbf{x}\|_2$ | (23) |
| $v_{\max,1} \triangleq \max_{\mathbf{x} \in \mathcal{B}} \|\hat{\mathbf{x}} - \mathbf{x}\|_2$ | (24) |
| $v_{\max,3} \triangleq \max_{\mathbf{x}, \mathbf{y} \in \mathcal{B}} \|\mathbf{x} - \mathbf{y}\|_2$ | (25) |
| **Upper Bounds:** | **Eqn.** |
| Bound1: | |
| $e \leqslant v_{\max,1} \leq \sqrt{v_{\mathrm{sdp}}\{\mathbf{A}_i, \mathbf{b}_i, c_i\}_{i=0}^N}$ | (27) |
| Bound2: | |
| $v_{\max,3} \leq 2R$ | (38) |
| Bound3 (Type 1): | |
| $v_{\max,3} \leq v_{\mathrm{socp}}$ | (45) |
| Bound3 (Type 2): | |
| $v_{\max,3} \leq v_{\mathrm{lp}}$ | (50) |

length of the intersection $\mathcal{B}$ is then upper bounded by

$$v_{\mathrm{lp}} = \sqrt{\sum_{i=1}^n (v_{\mathrm{lp},\ell})^2}. \tag{49}$$

Therefore, an upper bound on position error based on a bounding box approach is given by

$$v_{\max,3} \leq v_{\mathrm{lp}}. \tag{50}$$

It is clear that $v_{\mathrm{socp}} \leq v_{\mathrm{lp}}$.

Table I summarizes the various types of bounds derived in this study.

## VI. SIMULATION RESULTS

In this section we evaluate the validity of different upper bounds. We consider a $1000 \text{ m}^3$ cubic space for simulation. $N$ reference nodes are randomly distributed in the space. One target node is randomly placed inside the volume. To add measurement noise to actual distances between reference and target nodes, we use an exponential distribution defined as

$$f(\epsilon_i) = \begin{cases} \gamma e^{-\gamma \epsilon_i}, & \epsilon_i \geq 0 \\ 0, & \epsilon_i < 0. \end{cases}$$

The mean $1/\gamma$ is set to 1 m. The validity of exponential distribution, especially for NLOS conditions, has been justified in the literature, e.g., [11], [26], [35]. We study the POCS algorithm that always gives an estimate inside the intersection $\mathcal{B}$ in (17) . To solve the optimization problems formulated in this study, we use the *CVX* toolbox [36].

To evaluate the tightness of the bounds in Table I, we consider the normalized difference between a bound $v$ and the true error $e$, i.e., $(v - e)/e$. To illustrate how the tightness varies with, e.g., network deployment, measurement noise, estimator parameters, we study the cumulative distribution function (CDF)

$$P_v(x) = \Pr \left\{ \frac{v - e}{e} \leq x \right\},$$

where the randomness comes from selecting, e.g., the deployment in a random fashion. In the following, we will generate $e$ from POCS estimates. Since an estimate of the target position is available, we also consider the first upper bound for further comparisons. In all simulations, we generate 1000 random networks.

Fig. 8 shows the CDF of the normalized position error of an upper bound versus POCS position error for different number of reference nodes. As expected, the first upper bound shows better performance compared to the other bounds. For instance, Fig. 8(a) shows that in 80% of the cases, the first upper bound computed by the network consisting of five reference nodes is less than 2.3 times the actual position error (considering the normalized error $(v - e)/e$). This figure also shows that the upper bound 3 (Type 2) is the loosest bound. When the number of reference nodes increases, the upper bound 3 (Type 1) gets closer to the upper bound 2. Roughly speaking except for the upper bound 3 (Type 1), we can say that the behavior of other upper bounds (based on the normalized error $(v - e)/e$) does not change considerably with increasing the number of reference nodes. Fig. 8 also shows that the proposed bounds always are upper bounds (although not always tight).

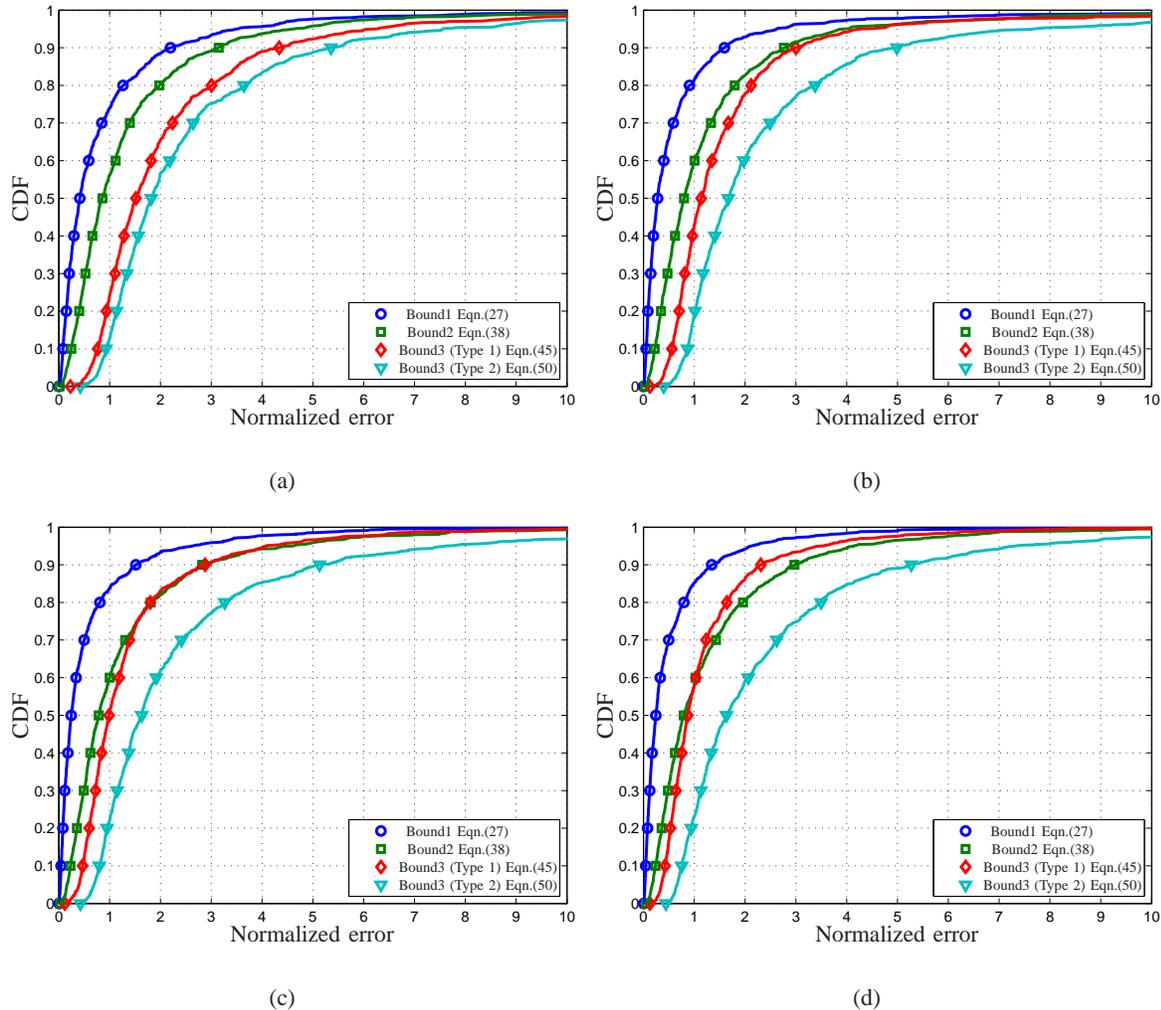

Fig. 8. Comparison between the CDF of normalized position error of upper bounds versus the POCS position error for, (a) 5 reference nodes, (b) 10 reference nodes, (c) 15 reference nodes, and (d) 20 reference nodes.

In the next simulation, we compare the upper bounds with the maximum position error. To compare four upper bounds, we again employ the POCS method. For every realization of the network, we run POCS for 200 random initializations and take the maximum position error. For every realization, the upper bound 1 corresponds to the maximum distance to the intersection for the estimate that gives the maximum POCS position error. Three other bounds are independent of the POCS estimate and they approximate the maximum length of the intersection area for every realization. Fig. 9 plots the four upper bounds against the maximum POCS position error. In Fig. 9(a), we plot the upper bound 1 and a lower bound on the maximum position error when an estimate is available. As seen, the maximum position error is bounded between the green and black curves, which defines an upper and a lower bounds on the maximum position error, respectively. These figures graphically show that the upper bound 1 is tighter than other bounds. They also show that the upper bound 3 (Type 2) is the loosest one.

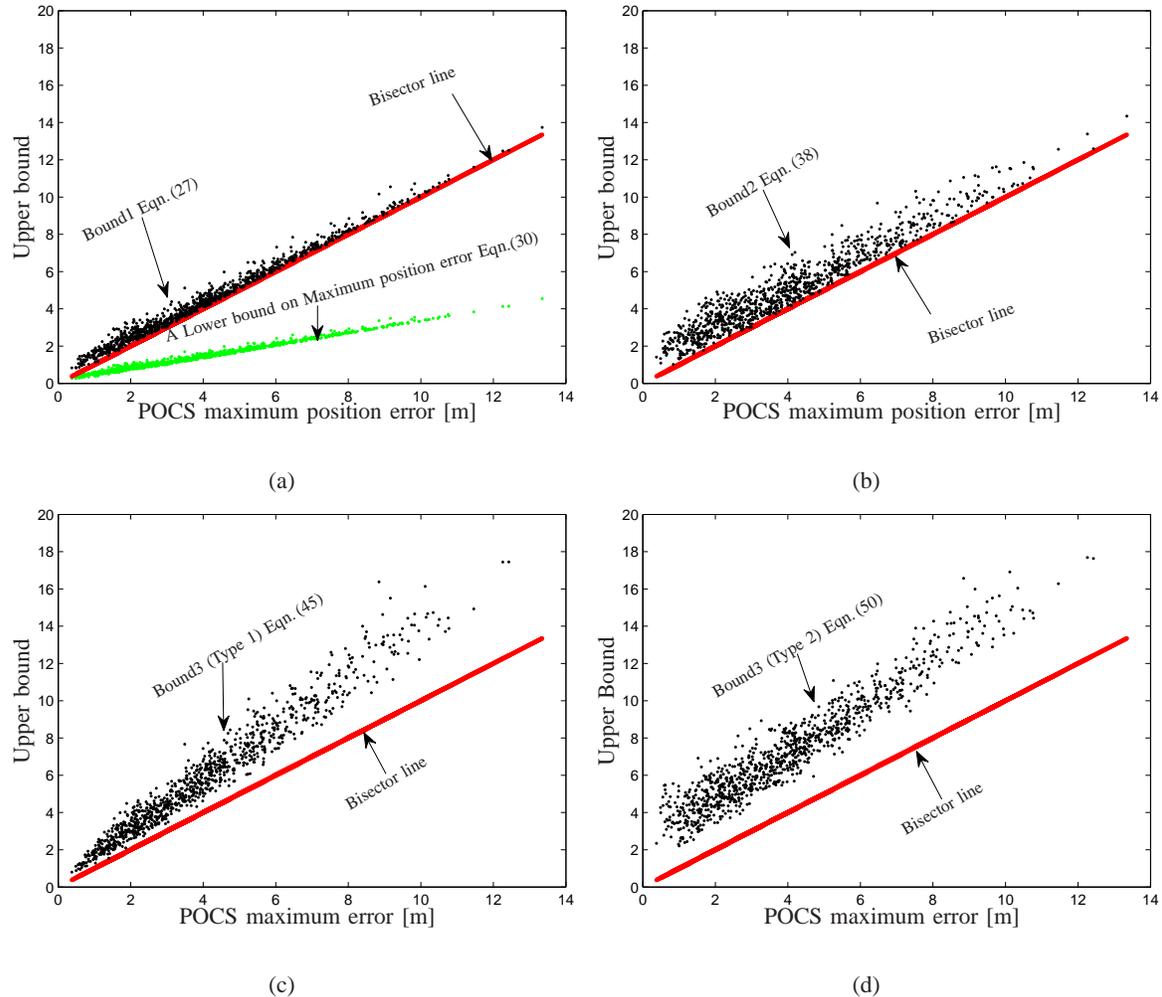

Fig. 9. Comparison between three upper bounds and the maximum position error of POCS for 15 reference nodes and 200 random initializations for every realization, (a) Bound1 is computed using the estimate that gives the maximum position error for POCS, (b) Bound2, (c) Bound3 (Type1), and (d) Bound3 (Type2).

In Fig. 10, we plot the CDF of the normalized position error of upper bounds versus the maximum POCS position error for different number of reference nodes. Roughly speaking, in more than 90% of cases the upper bound 1 is equal or less than 1.5 times the maximum POCS position error for different number of reference nodes. Again, we see that the upper bound 1 is the tightest and the upper bound 3 (Type 2) is the loosest one. It is seen that when the number of reference nodes increases to 15, the upper bound 2 in 80% of cases is tighter that the upper bound 3 (Type 1).

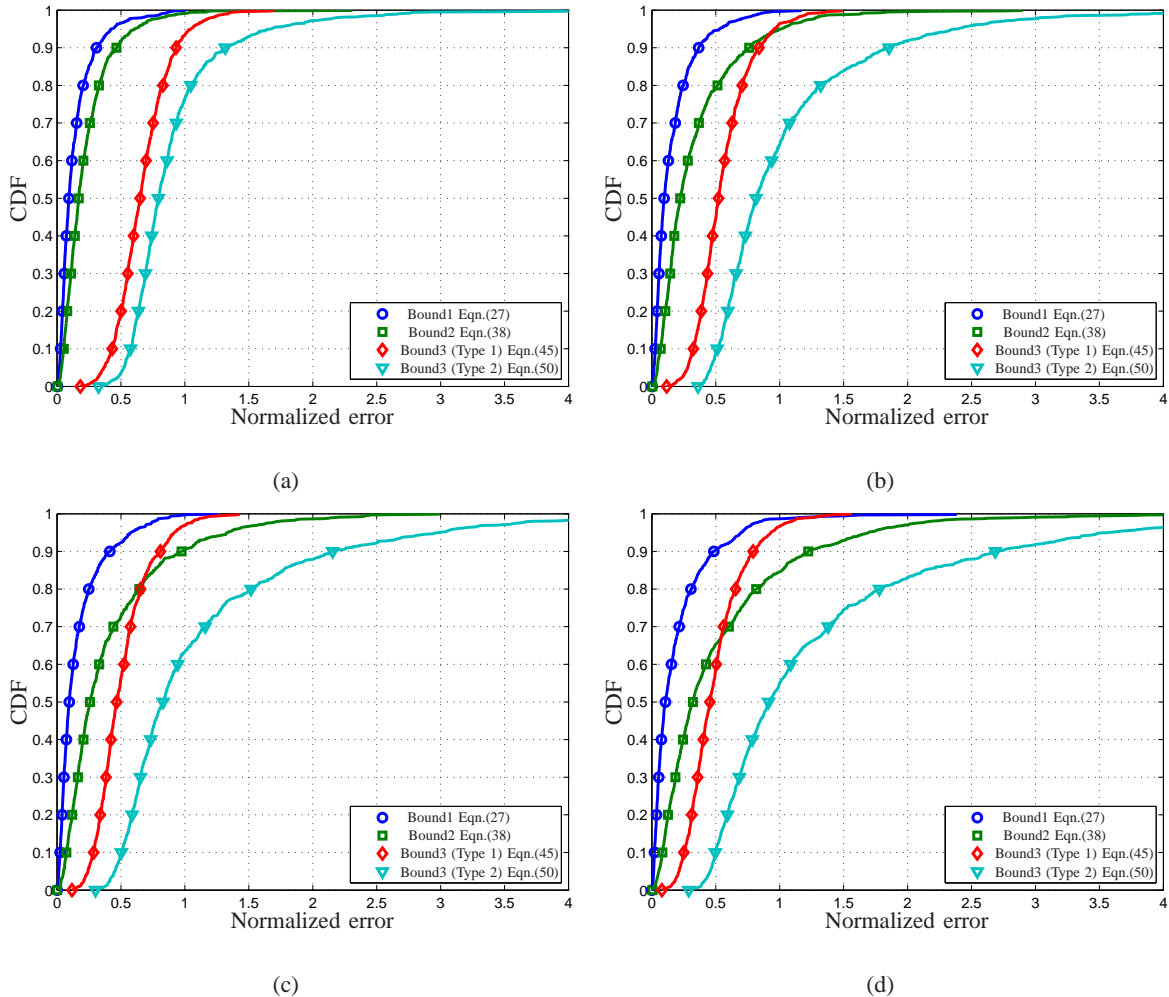

Fig. 10. Comparison between the CDF of normalized error of different upper bounds versus the maximum position error of POCS for, (a) 5 reference nodes, (b) 10 reference nodes, (c) 15 reference nodes, and (d) 20 reference nodes.

## VII. Conclusions

In this paper we have formulated a number of upper bounds on the realization of the positioning error, i.e., the error which is produced by an estimator, or a class of estimators, given a certain realization of the measurement, $\mathbf{m}$. The bound defined in (8) can be computed by finding the largest distance between a point in the set $\mathcal{X}(\mathbf{m})$, i.e., the set of all possible positions of the unknown node, conditioned on the observation $\mathbf{m}$, and the estimate $\hat{\mathbf{x}}(\mathbf{m}, \mathbf{f})$. (Recall that $\mathbf{f}$ contains the estimation algorithm parameters, e.g., initialization, that determines how $\mathbf{m}$ is mapped to the position estimate.) Similarly, the bound in (10) can be computed as the largest distance between a point in $\mathcal{X}(\mathbf{m})$ and a point in $\hat{\mathcal{X}}(\mathbf{m}) = \{\hat{\mathbf{x}}(\mathbf{m}, \mathbf{f}) : \mathbf{f} \in \mathcal{F}\}$, i.e., the set of all possible estimates in the class of estimators defined by $\mathcal{F}$. Hence, the bounds are nontrivial (i.e., finite) only if the measurement implies that the above-mentioned sets are of finite lengths. Moreover, it is, in general, not clear if the bounds can be computed with reasonable complexity.

However, we have showed that we can indeed compute nontrivial bounds in an efficient manner for the special, but interesting, case when $\mathbf{m}$ consists of positively biased distances estimates between a number of reference (anchors nodes) at a-priori known positions and a target node (at an unknown position). We note that non-negative distance errors are likely to occur in non-line-of-sight environments. For this special case, the target node is constrained to be in the intersection $\mathcal{B}$ of a number of balls, $\mathcal{B}_i$, $i = 1, 2, \ldots, N$, which are centered around the reference nodes and whose radii are given by the observed distance estimates. That is, in this special case, $\mathcal{X}(\mathbf{m}) = \mathcal{B}$. An efficient algorithm, (27), can then be found by relaxing the original bound (24) into a convex optimization problem using SDP techniques.

Moreover, if we use a POCS algorithm to estimate the target node position, we know that $\hat{\mathcal{X}}(\mathbf{m}) = \mathcal{B}$, i.e., the estimate will be in $\mathcal{B}$. Hence, the bound (8) simplifies to (25). To arrive at bounds that can be efficiently computed, we formulate three upper bounds of (25) in (38), (45), and (50). The bound (38) is based on SDP relaxation, the bound (45) by replacing $\mathcal{B}$ with its bounding box in (25), and the bound (50) by replacing $\mathcal{B}_i$ with their bounding boxes in (17). Simulation results based on the POCS estimate for different situations show that the proposed upper bounds provide reasonably tight bounds. As expected from the theoretical part and confirmed by the simulation results, for the POCS estimate the first bound in (27) is the tightest bound among different upper bounds formulated in this paper. The numerical results also show that the behavior of different bounds, except the one in (45), based on the normalized error does not considerably change with node density. It is also concluded from both theoretical aspects and simulation results that the bounds (38) and (45) are tighter than the one in (50).

Finally, it is clear that it is very valuable if we, in a practical situation, can append an estimated position with an upper bound of the position error. This is much stronger than saying something about the statistics of the position error (e.g., the mean squared error). The methods developed in this paper provides tools for bounding the position error, albeit in somewhat limited situations, i.e., when $\mathcal{X}(\mathbf{m})$ has finite length. There are practical situations where this is a valid assumption, but also cases when it is not.

## VIII. ACKNOWLEDGMENT

Authors would like to thank Prof. Stephen P. Boyd for comments on the optimization problems considered in this paper. They also would like to thanks Dr. Sinan Gezici for comments on the paper.